\documentclass[11pt]{article}
\pdfoutput=1
\usepackage[latin1]{inputenc}
\usepackage{amsmath}
\usepackage{amsfonts}
\usepackage{amssymb}
\usepackage{pstricks}
\usepackage{amsthm}
\usepackage{mathrsfs}
\usepackage{amssymb}
\usepackage{cancel}
\usepackage{slashed}
\usepackage{graphicx}
\usepackage[font=small,labelfont=bf]{caption}
\usepackage{float}
\usepackage{csquotes}
\usepackage[bottom]{footmisc} 
\usepackage{tikz}
\usepackage{setspace}
\usepackage{bbding}
\usepackage{cite}
\usetikzlibrary{matrix,arrows,decorations.pathmorphing}
\usepackage{framed}
\usepackage{color}
\usepackage{wrapfig}
\definecolor{shadecolor}{RGB}{224,238,238}
\makeatletter

\@addtoreset{equation}{section}
\makeatother

\def\lsim{\;\raise0.3ex\hbox{$<$\kern-0.75em\raise-1.1ex\hbox{$\sim$}}\;}
\def\gsim{\;\raise0.3ex\hbox{$>$\kern-0.75em\raise-1.1ex\hbox{$\sim$}}\;}
\def\beq{\begin{equation}}   \def\eeq{\end{equation}}
\def\ba{\begin{array}}       \def\ea{\end{array}}
\def\bea{\begin{eqnarray}}   \def\eea{\end{eqnarray}}

\def\nl{\newline}

\def\noi{\noindent}

\theoremstyle{definition} 
\date{\today}

\begin{document}

\begin{titlepage}
\begin{flushright}
LPT Orsay 18-74\\
LUPM 18-026
\end{flushright}


\begin{center}

\begin{doublespace}

\vspace{1cm}
{\Large\bf The higgsino-singlino sector of the NMSSM: Combined
constraints from dark matter and the LHC} 
\vspace{2cm}

{\bf{Ulrich Ellwanger$^{a}$ and Cyril
Hugonie$^b$}}\\
\vspace{1cm}
{\it  $^a$ Laboratoire de Physique Th\'eorique, UMR 8627, CNRS, Universit\'e de Paris-Sud, Universit\'e
Paris-Saclay, 91405 Orsay, France\\
\it $^b$  LUPM, UMR 5299, CNRS, Universit\'e de Montpellier, 34095
Montpellier, France }

\end{doublespace}

\end{center}
\vspace*{2cm}
\begin{abstract}
A light singlino is a promising candidate for dark matter, and a light higgsino
is natural in the parameter space of the NMSSM. We study the combined constraints
on this scenario resulting from the dark matter relic density, the most recent
results from direct detection experiments, LEP and the LHC. In particular
limits from a recent search for electroweak production of charginos and neutralinos
at $\sqrt{s}=13$~TeV after 35.9~fb$^{-1}$ by CMS and constraints on spin-independent
dark matter-nucleon cross sections from XENON1T after one tonne$\times$year
exposure are considered.
We find that scenarios with higgsino masses below $\sim 250$~GeV as well as
singlino masses below $\sim 100$~GeV are strongly constrained depending,
however, on assumptions on the bino mass parameter $M_1$.
Benchmark points and branching fractions for future searches at the LHC are proposed.
\end{abstract}

\end{titlepage}

\newpage
\section{Introduction}
\label{sec:intro}

One of the promising aspects of supersymmetric extensions of the Standard Model
is the natural presence of dark matter if $R$-parity is unbroken and the
lightest supersymmetric particle (LSP) is neutral. Candidates for such LSPs
are the neutral electroweak gauginos (bino and wino), the neutral fermionic
partners of Higgs doublets (higgsinos) and, in the Next-to-Minimal Supersymmetric
Standard Model (NMSSM), the fermionic partner of a scalar singlet (singlino).

Experiments searching for interactions of dark matter with standard matter have
made considerable progress in the last years. The absence of signals has lead
to upper bounds on dark matter-standard matter interaction cross sections, both
in spin-independent (SI) and in spin-dependent (SD) channels. Such upper bounds
have been obtained recently in SI channels by the PandaX-II \cite{Tan:2016zwf},
LUX \cite{Akerib:2016vxi} and XENON1T collaborations \cite{Aprile:2017iyp,Aprile:2018dbl},
and in SD channels by the PICO-2L \cite{Amole:2016pye}, LUX \cite{Akerib:2016lao}
and PandaX-II \cite{Fu:2016ega} collaborations.

Since the higgsino mass parameter $\mu$ is supersymmetric, a large value
$|\mu| \gg M_Z$ would generate a ``little fine-tuning problem'': The potential for
the scalar Higgs doublets $H_u$ and $H_d$ contains positive mass terms
$|\mu|^2$, but it must be unstable at the origin to trigger $SU(2)\times U(1)$
symmetry breaking. Hence at least one of the positive mass terms $|\mu|^2$
must be cancelled by a negative soft supersymmetry breaking mass term. In
order to generate Higgs vacuum expectation values (vevs) of ${\cal O} (M_Z)\ll |\mu|$,
this cancellation would have to be fine-tuned if $|\mu| \gg M_Z$.
Hence a higgsino mass parameter $|\mu|$ not far above $M_Z$ is natural.

However, assuming a standard thermal history of the universe and that the lightest
supersymmetric particle (LSP) accounts for the complete relic density
$\Omega_{DM} h^2 \sim 0.1187$ in agreement with WMAP/Planck \cite{Hinshaw:2012aka,Ade:2013zuv},
mostly higgsino-like dark matter is strongly constrained. In order to avoid
a too large annihilation rate its mass is $\gsim 1$~TeV according to
\cite{Baer:2016ucr,Athron:2017qdc,Profumo:2017ntc,Roszkowski:2017nbc,
Bagnaschi:2017tru,Kowalska:2018toh,Baer:2018rhs} unless 
scalar top squarks (stops) are very heavy \cite{Huang:2017kdh,Badziak:2017the}
or $\mu < 0$ \cite{Abdughani:2017dqs}.

In the NMSSM an effective $\mu$ parameter is generated by the vev of a
scalar singlet $S$, $\mu_\text{eff}=\lambda \left< S\right>$ \cite{Maniatis:2009re,
Ellwanger:2009dp}. The fermionic partner of $S$, the singlino, is a promising
dark matter candidate \cite{Cerdeno:2004xw,
Belanger:2005kh,Cerdeno:2007sn,Barger:2007nv,Belanger:2008nt,Vasquez:2010ru,
Perelstein:2012qg,Kozaczuk:2013spa,Cao:2013mqa,Kim:2014noa,
Ellwanger:2014dfa,Ishikawa:2014owa,
Han:2014nba,Cheung:2014lqa,Huang:2014cla,Cahill-Rowley:2014ora,
Guo:2014gra,Cao:2014efa,Bi:2015qva,Cao:2015loa,Butter:2015fqa,
Gherghetta:2015ysa,Han:2015zba,Potter:2015wsa,Barducci:2015zna,Enberg:2015qwa,
Badziak:2015exr,Xiang:2016ndq,Cao:2016nix,Cao:2016cnv,Ellwanger:2016sur,Beskidt:2017xsd,
Mou:2017sjf,Badziak:2017uto,Baum:2017enm} which can account for the observed relic density
and have sufficiently small dark matter-standard matter interaction
cross sections, see section 3. The singlino can be very light; then $\mu_\text{eff}$
can be small as well (provided the higgsinos remain heavier than the singlino)
solving this ``little fine-tuning problem''. Only a lower bound $|\mu_\text{eff}|\gsim
100$~GeV originates from the non-observation of a charged higgsino at LEP.
Hence the ``light higgsino-singlino scenario'' in the NMSSM is quite attractive
\cite{Ellwanger:2013rsa,Ellwanger:2014hia,Dutta:2014hma,Jeong:2014xaa,
Allanach:2015cia,Kim:2015dpa,Akula:2017yfr} (although an explanation of the
galactic center gamma-ray excess seems to be difficult \cite{Shang:2018dja}).

Higgsinos (and winos) can be produced at colliders through electroweak processes.
Denoting the lightest chargino by $\chi^\pm_1$ and the neutralinos by $\chi^0_i$
(ordered in mass) their typical decays are $\chi^\pm_1 \to W^{\pm (*)} + \chi^0_1$,
$\chi^0_i \to Z^{(*)} + \chi^0_1$ or $\chi^0_i \to H^{(*)} + \chi^0_1$
($i > 1$) where $H$
can correspond to the SM-like Higgs boson $H_{SM}$ or, notably in the NMSSM, to a lighter
mostly singlet-like CP-even or CP-odd scalar.

The most promising search channel is then $pp \to W^{\pm *} \to \chi^\pm_1 + \chi^0_i$
with $E_T^{miss}$ and three leptons from leptonic decays of $W^{\pm (*)}$
and $Z^{(*)}$ \cite{Baer:1985at,Baer:1986vf,Baer:1986dv}.
At the LHC with 13~TeV c.m. energy and $\sim 36$ fb$^{-1}$ of integrated luminosity
these signatures -- including hadronic decays of $W$ and $Z$, and
$\chi^0_i \to H_{SM} + \chi^0_1$ --
have been searched for by ATLAS \cite{Aaboud:2017leg,Aaboud:2018jiw,Aaboud:2018sua,Aaboud:2018htj}
and CMS \cite{Sirunyan:2018ubx,Sirunyan:2017lae}. No significant excesses have
been observed which leads to upper bounds on $\chi^\pm_1 + \chi^0_i$ production
cross sections.

It is the aim of the present paper to study the combined constraints on the
light higgsino-singlino scenario in the NMSSM from the dark matter relic density,
spin-dependent and spin-independent direct detection experiments and from neutralino/chargino
searches at the LHC at 36~fb$^{-1}$.

We consider two ${\mathbb Z}_3$-invariant versions of the NMSSM:
the phenomenological model (pNMSSM) with arbitrary parameters at the weak scale,
and the Non Universal Higgs model (NUH-NMSSM) with universal gaugino masses
$M_{1/2}$, universal sfermion masses masses $m_0$ and universal sfermion trilinear soft
terms $A_0$ but arbitrary Higgs soft masses $m_{H_u}$, $m_{H_d}$ and $m_S$ as well as Higgs
trilinear soft terms $A_\lambda$ and $A_\kappa$.


We assume heavy squark, slepton and gluino masses well above $1$~TeV;
the squark and gluino masses have no impact on the light higgsino-singlino scenario.
On the other hand bino and/or wino masses $M_1$ and $M_2$, respectively, can affect the
production cross sections and branching fractions of the (mostly) higgsino-like
chargino and neutralinos through mixing (see below).
We consider two versions of the pNMSSM, both with $M_2=600$~GeV but one with
$M_1=300$~GeV, another one with $M_1$ arbitrary.
In the NUH-NMSSM a (conservative) lower bound $M_3 \gsim 1.6$~TeV on
the gluino mass term at the weak scale implies $M_{1/2}\gsim 440$~GeV and hence
$M_1\gsim 220$~GeV, $M_2\gsim 370$~GeV at the weak scale. $M_1$, $M_2$ have some impact on
$A_\lambda$ and $A_\kappa$ at the weak scale via the renormalization group
equations, hence the NUH-NMSSM implies some correlations among
the parameters of the pNMSSM. The scans
over the parameter spaces are performed with help of the public code
{\sf NMSSMTools}~\cite{Ellwanger:2004xm,Ellwanger:2005dv,Das:2011dg}.
The dark matter relic density and the spin-dependent and spin-independent
direct detection cross sections are computed with help of
{\sf micr\-OMEGAS\_3}~\cite{Belanger:2013oya}.

In the next Section we review the relevant parts of the neutralino sector of the
NMSSM, and in Section~3 the impact of a viable dark matter relic density and
bounds from direct dark matter detection experiments. In Section~4 we discuss
the implementation of bounds from searches for $Z+W+E_T^{miss}$
at the LHC, in Section~5 the resulting
constraints in the plane $M_{\chi^0_1}$ vs. $M_{\chi^\pm_1}$. In Section~6 we
propose benchmark points and planes, and discuss realistic branching fractions to be used
for future searches; Section~7 is devoted to a summary.

\section{The neutralino sector of the NMSSM}

We consider the ${\mathbb Z}_3$ invariant NMSSM with the superpotential
\beq\label{eq:2.1}
W_\text{NMSSM} = \lambda \hat S \hat H_u\cdot \hat H_d + \frac{\kappa}{3} 
\hat S^3 +\dots
\eeq
where the dots denote the Yukawa couplings of the superfields $\hat H_u$ and $\hat H_d$
to the quarks and leptons as in the MSSM. Once the scalar component of the superfield
$\hat S$ develops a vev $\left< S\right>\equiv s$, the first term in
$W_\text{NMSSM}$ generates an effective $\mu$-term with
\beq\label{eq:2.2}
\mu_\mathrm{eff}=\lambda\, s\; .
\eeq
Subsequently the index $_\mathrm{eff}$ of $\mu$ will be omitted for
simplicity. $\mu$ generates Dirac mass terms for the charged and
neutral SU(2) doublet higgsinos $\psi_u$ and $\psi_d$.

In the ``decoupling'' limit $\lambda, \kappa \to 0$ all components of the superfield
$\hat S$ decouple from all components of $\hat H_u,\ \hat H_d$ and the matter
superfields. However, since $s\sim M_{Susy}/\kappa$ where $M_{Susy}$ denotes
the scale of soft Susy breaking masses and trilinear couplings, 
$\mu_\mathrm{eff}$ remains of ${\cal O}(M_{Susy})$ in the decoupling limit
provided $\lambda/\kappa\sim {\cal O}(1)$.

Including bino ($\widetilde{B}$) masses $M_1$ and wino ($\widetilde{W}^3$) masses
$M_2$, the symmetric $5 \times 5$ neutralino mass matrix ${\cal M}_0$
in the basis $\psi^0 = (-i\widetilde{B} ,
-i\widetilde{W}^3, \psi_d^0, \psi_u^0, \psi_S)$
is given by \cite{Ellwanger:2009dp}
\beq\label{eq:2.3}
{\cal M}_0 =
\left( \ba{ccccc}
M_1 & 0 & -\frac{g_1 v_d}{\sqrt{2}} & \frac{g_1 v_u}{\sqrt{2}} & 0 \\
& M_2 & \frac{g_2 v_d}{\sqrt{2}} & -\frac{g_2 v_u}{\sqrt{2}} & 0 \\
& & 0 & -\mu & -\lambda v_u \\
& & & 0 & -\lambda v_d \\
& & & & 2\kappa s
\ea \right)
\eeq
where $v_u^2 + v_d^2 =v^2 \simeq (174\ \text{GeV})^2$ and $\frac{v_u}{v_d}=\tan\beta$.
The eigenstates of ${\cal M}_0$ are denoted by $\chi_i^0$, $i=1...5$ ordered in mass.
Henceforth the LSP is identified with $\chi_1^0$.

Another important r\^ole will be played by the singlet-like scalar and pseudoscalar
Higgs masses. The CP-even sector comprises three physical states which are linear
combinations of the real components $(H_{dR}, H_{uR}, S_R)$.
The (3,3) element of the $3 \times 3$ CP-even mass matrix ${\cal
M}_S^2$ reads in this basis
\beq\label{eq:2.4}
{\cal M}_{S,33}^2 \equiv M_{S_R,S_R}^2  =  \lambda A_\lambda \frac{v_u v_d}{s}
+ \kappa s (A_\kappa + 4\kappa s)\; ;
\eeq
up to mixing it corresponds to the mass squared of the mostly singlet-like eigenstate.
Another eigenstate must correspond to a Standard~Model-like Higgs boson $H_{SM}$ with its mass
$\sim 125$~GeV and nearly Standard~Model-like couplings to quarks, leptons and gauge
bosons. A third MSSM-like eigenstate has a mass of about
$\displaystyle{2\frac{\mu(A_\lambda+\kappa s)}{\sin 2\beta}}$. In the regions of the parameter space
of interest here we always find
that the mostly singlet-like eigenstate is the lightest CP-even scalar $H_1$, the
Standard-Model-like Higgs boson $H_{SM}$ is the second lightest CP-even scalar $H_2$, and
the MSSM-like state is the third CP-even scalar $H_3$.

The CP-odd sector consists in linear combinations of the imaginary components
$(H_{dI}, H_{uI}, S_I)$. The (3,3) element of the $3 \times 3$ CP-odd mass matrix ${\cal
M}_P^2$ reads in this basis
\beq\label{eq:2.5}
{\cal M}_{P,33}^2 \equiv M_{S_I,S_I}^2  =  \lambda (A_\lambda+4\kappa s)\frac{v_u
v_d}{s} -3\kappa A_\kappa s\; ;
\eeq
again it corresponds essentially to the mass squared of the mostly singlet-like eigenstate.
Other eigenstates are the electroweak Goldstone boson, and an MSSM-like eigenstate again with a mass of
about $\displaystyle{2\frac{\mu(A_\lambda+\kappa s)}{\sin 2\beta}}$. The masses
of the MSSM-like Higgs bosons are bounded from below by constraints
from $b\to s + \gamma$ on the charged Higgs boson whose mass is similar to the ones of the 
CP-even and CP-odd neutral scalars, and by direct searches \cite{Aaboud:2017sjh,Sirunyan:2018zut}.
Subsequently the lightest mostly singlet-like CP-odd eigenstate will be denoted by $A_1$.

From eqs.~\eqref{eq:2.3}-\eqref{eq:2.5} one can derive the sum rule \cite{Das:2012rr}
\beq\label{eq:2.6}
M_{\psi_S,\psi_S}^2\equiv 4\kappa^2 s^2 = M_{S_R,S_R}^2 + \frac{1}{3} M_{S_I,S_I}^2 -\frac{4}{3}
v_u v_d\left(\lambda^2\frac{A_\lambda}{\mu}+\kappa\right)
\eeq
which relates, up to modifications by mixing, the singlet-like neutralino, CP-even and CP-odd Higgs
masses. In the decoupling limit, or for sizeable $\tan\beta$ (i.e. small $v_d$) and not too large
$A_\lambda$ and Yukawa couplings $\lambda$ and $\kappa$, the last term in eq.~\eqref{eq:2.6} is negligible.

\section{Dark matter relic density and direct detection}

As sketched in the Introduction, under the assumption of a standard thermal history
of the universe and $|\mu|$ well below 1~TeV the mostly singlet-like neutralino
$\psi_S$ remains practically the only viable candidate for dark matter.
Its annihilation rate must be sufficiently large such that its relic
density today complies with the WMAP/Planck value $\Omega_{DM}h^2\simeq 0.1187$
\cite{Hinshaw:2012aka,Ade:2013zuv}. Various processes can give rise to a large enough
annihilation cross section:

\noi a) Annihilation via a pseudoscalar in the s-channel. At least for singlino masses $M_{\chi_1^0}$ below
$\mu$ as assumed here this pseudoscalar is the singlet-like $A_1$
with its mass given in eq.~\eqref{eq:2.5} (up to a small shift through mixing).
$M_{A_1}$ should be about $2\times M_{\chi_1^0}$ such
that the annihilation cross section is enhanced by the s-channel pole (depending on $\kappa$
and the mixing of $A_1$ with the MSSM-like SU(2)-doublet pseudoscalar which induces its couplings
to quarks and leptons). For $M_{\chi_1^0}$ above $\approx 100$~GeV, $M_{A_1}$ and hence the
width of $A_1$ increase and $M_{A_1}$ can be smaller than $2\times M_{\chi_1^0}$
allowing for LSP annihilation via $A_1^* \to A_1+H_1$ provided $M_{H_1}$ is small enough.
For $M_{\chi_1^0}$ above $ m_{top}$ the annihilation via $A_1^* \to t\bar{t}$ becomes
possible.

\noi b) Annihilation via the $Z$ boson or the Standard~Model-like Higgs boson $H_{SM}$ in the
s-channel if the singlino mass is about half the $Z$ or $H_{SM}$ mass.

\noi c) Annihilation via a far off-shell $Z$ boson into $t\bar{t}$ if $M_{\chi_1^0}> m_{top}$.

\noi d) Annihilation into a pair of $W/Z$ bosons via (higgsino-like) chargino/neutralino exchange
in the t-channel. This t-channel process is strong enough to be dominant only for singlino
masses above $\sim 100$~GeV.

\noi e) 
Coannihilation with higgsinos becomes relevant for $M_{\chi_1^0}\sim \mu$.

\noi f) 
Coannihilation with staus $\tilde{\tau}$ becomes relevant for $M_{\chi_1^0}\sim M_{\tilde{\tau}}$.

In the case of annihilation via $A_1 \sim S_I$ in the s-channel with a pseudoscalar mass about twice
the mass of the singlino eq.~\eqref{eq:2.6} leads to
\beq\label{eq:2.7}
M_{S_R,S_R}^2=
-\frac{1}{3}M_{\psi_S,\psi_S}^2+\frac{4}{3}
v_u v_d\left(\lambda^2\frac{A_\lambda}{\mu}+\kappa\right) 
\eeq
implying an intolerable negative CP-even scalar mass squared if off-diagonal terms in the
mass matrices and the terms $\sim v_u v_d$ are neglected, unavoidably
in the decoupling limit $\lambda, \kappa \to 0$.
Hence most scenarios with singlino annihilation via a $A_1$ in the s-channel are not
compatible with the decoupling limit; in such scenarios we found $\lambda \gsim 0.2$ 
(see Figure~4 below). Only if the mostly singlino-like LSP is lighter than $\sim 20$~GeV,
smaller values of $\lambda$ can suffice to generate the required mixing in the
neutralino mass matrix~\eqref{eq:2.3} in order to avoid $M_{H_1}^2 < 0$.
Still $M_{H_1}$ tends to be small, and $A_\lambda$ to be
large in this scenario.

The most recent and most stringent constraints on dark matter detection direct
cross sections are from Xenon1T \cite{Aprile:2018dbl} (spin-independent) and
PandaX-II \cite{Fu:2016ega} (spin-dependent). These constraints are still very
weak for LSP masses below a few GeV, but affect the present scenario for
LSP masses above $\sim 5$~GeV.

Contributions to spin-independent singlino-nucleon cross sections originate from exchanges
of the SM-like Higgs boson and the mostly singlet-like $H_1$ in the t-channel.
The coupling of the latter to nucleons is even smaller than the one of the SM-like Higgs boson,
but $H_1$ is typically much lighter (see above) and has larger
couplings to the mostly singlino-like LSP. The contributions from the SM-like Higgs
boson and $H_1$ interfere negatively, and can thus reduce the
spin-independent singlino-nucleon cross section below the neutrino floor
 \cite{Ellwanger:2014dfa,Badziak:2015exr}.

At first sight the upper bounds on spin-dependent singlino-nucleon cross sections are about
five orders of magnitude weaker. However, contributions to spin-dependent LSP-nucleon
cross sections originate from $Z$-exchange in the t-channel, and $Z$-nucleon couplings are
much larger than Higgs-nucleon couplings.
A $\chi^0_1-\chi^0_1-Z$ coupling originates from higgsino components of
the mostly singlino-like $\chi^0_1$ induced by mixing $\sim \lambda$. For a light
mostly singlino-like LSP and not too small $\tan\beta$ this higgsino component is
$\approx \lambda \times (174\ \text{GeV}/\mu)$,
which is thus bounded from above by upper limits on the spin-dependent singlino-nucleon cross section
depending on $M_{\chi^0_1}$ from PandaX-II. (For LSP masses below a few GeV where
constraints from direct detection are weak, the higgsino component of the LSP is still bounded from
above by its contribution to the invisible $Z$ width.)

\section{Constraints from searches for charginos and neutralinos at the LHC}

As stated in the Introduction, the most stringent LHC bounds originate from
searches for  $pp  \to \chi^\pm_1 + \chi^0_i
\to W^\pm + Z + E_T^{miss}$, dominantly from trileptons. The absence of significant excesses
can be interpreted as upper bounds on production cross sections times branching
fractions of charginos and neutralinos within simplified models. In  Figs.~7 and 8a in
\cite{Sirunyan:2018ubx}, upper bounds on production cross sections of
charginos $\chi^\pm_1$ and neutralinos $\chi^0_2$ are given as function
of their mass (assumed to be degenerate), and the mass of $\chi^0_1$.
We used the data in root format underlying these Figures. 
In the same Figures, limits on $M_{\chi^\pm_1}$ as function of $M_{\chi^0_1}$ are given assuming
100\% branching fractions for the decays $\chi^\pm_1 \to W^{\pm (*)} + \chi^0_1$
and $\chi^0_2 \to Z^{(*)} + \chi^0_1$, and assuming production cross sections for
wino-like charginos $\chi^\pm_1$ and neutralinos $\chi^0_2$.

As a first step we re-interprete the data as upper bounds on the sum of production cross sections
times branching fractions of
pure higgsino-like charginos $\chi^\pm$ and higgsino-like neutralinos $\chi^0_2$ and $\chi^0_3$
assuming $M_{\chi^\pm} = M_{\chi^0_2} = M_{\chi^0_3}$, as function of $M_{\chi^\pm}$ 
and the mass of $\chi^0_1$. Note that the production cross section for
higgsinos is only half the one for winos despite the sum over $\chi^0_2$ and $\chi^0_3$.
(Higgsino pair production has also been considered by ATLAS and CMS in
\cite{Aaboud:2017leg,Aaboud:2018jiw,Sirunyan:2018ubx,Sirunyan:2017lae}. The assumed
higgsino decays there differ, however, significantly from the scenario considered here
where the higgsino decays are similar to the ones assumed for winos in
\cite{Aaboud:2017leg,Aaboud:2018jiw,Sirunyan:2018ubx,Sirunyan:2017lae}.)

It is instructive to compare this upper bound for a light LSP
of mass $M_{\chi^0_1}=5$~GeV
to the production cross sections of pure higgsino-like charginos and neutralinos,
assuming a common higgsino mass. The upper limits from CMS  in
\cite{Sirunyan:2018ubx} are shown as function of a common chargino/neutralino mass as a red line in Fig.~1;
the zig-zag behaviour (present in the root files) seems to originate from the combination of
different signal regions. The sum of production cross sections for
pure higgsino-like charginos and both higgsino-like neutralinos from the
LHC SUSY Cross Section Working Group twiki page \cite{HiggsinoXS} is shown as a blue line.

\begin{figure}!t
\begin{center}
\includegraphics[width=11cm]{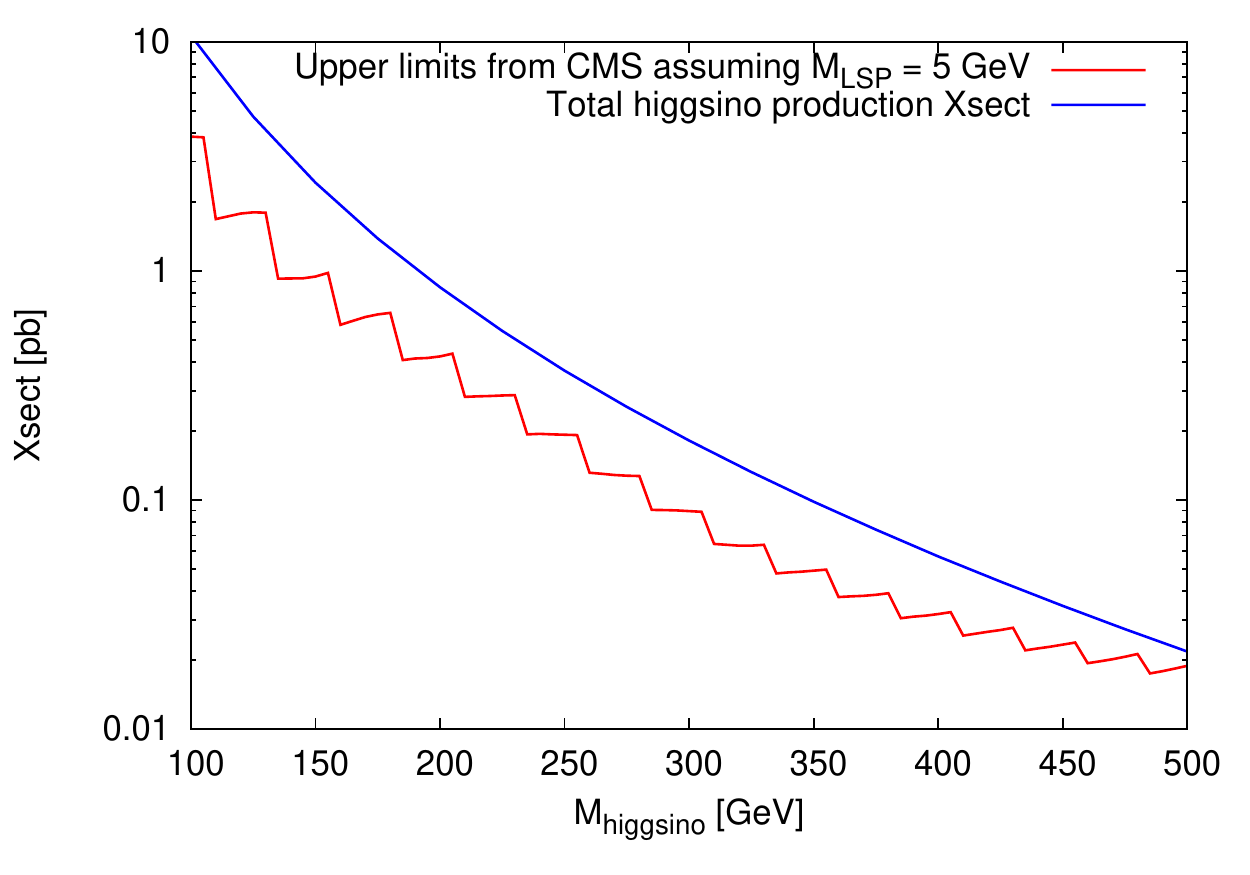}
\end{center}
\caption{Red line: Upper limits on the production cross section times
branching fraction for $pp  \to \chi^\pm_1 + \chi^0_i \to W^\pm + Z + E_T^{miss}$,
assuming $M_{\chi^0_1}=5$~GeV and 100\% branching fractions for the decays
$\chi^\pm_1 \to W^{\pm (*)} + \chi^0_1$ and $\chi^0_2 \to Z^{(*)} + \chi^0_1$,
from \cite{Sirunyan:2018ubx}. Blue line: Production cross sections from
the LHC SUSY Cross Section Working Group twiki page for
degenerate pure higgsino-like charginos and a pair of higgsino-like neutralinos.}
\end{figure}

One observes that the limits are compatible with the production of
pure higgsino-like charginos and higgsino-neutralinos only if their branching
fractions into the considered final states are below 100\%, the smaller for
smaller $\mu$.  Hence we have to shed some light on the branching fractions of
the higgsino-like charginos and neutralinos in the present
scenario.

First, unless the stau's or the bino are light, the chargino decay
$\chi^\pm_1 \to W^{\pm (*)} + \chi^0_1$
is the only possible decay and has a branching ratio of 100\%. The possible decays
of the neutral higgsinos $\chi^0_{2,3}$ are more involved: The couplings allowing
for $\chi^0_{2,3} \to Z^{(*)} + \chi^0_1$ originate from the higgsino components
$\sim \lambda$ of $\chi^0_1$, hence these partial widths are of ${\cal O}(\lambda^2)$
in the decoupling limit. Yukawa couplings for the decays $\chi^0_{2,3} \to H_{SM} + \chi^0_1$
do not require mixing, but are induced by the term $\lambda \hat S \hat H_u\cdot \hat H_d$
in the superpotential. Consequently these partial widths are equally of ${\cal O}(\lambda^2)$.

We find that, if both decays into $Z$ and $H_{SM}$ are kinematically allowed and after averaging
over $\chi^0_{2}$ and $\chi^0_{3}$, $BR(\chi^0_{2,3} \to Z^{(*)} + \chi^0_1)\approx 50-70\%$,
and $BR(\chi^0_{2,3} \to H_{SM} + \chi^0_1)\approx 30-50\%$. Upper bounds on the latter
process have also been provided by CMS in \cite{Sirunyan:2018ubx}, but the corresponding
limits are considerably weaker than the ones from $\chi^0_{2,3} \to Z^{(*)} + \chi^0_1$
(to which we confine ourselves).
Hence an enhanced $BR(\chi^0_{2,3} \to H_{SM} + \chi^0_1)$ alleviates the constraints
from \cite{Sirunyan:2018ubx}, and this happens notably for a bino mass term $M_1$ close to $\mu$ which
induces bino components of the $\chi^0_{2,3}$ through mixing. These bino components reduce 
the couplings of $\chi^0_{2,3}$ to $Z$ (but not to $H_{SM}$) and reduce their production cross sections. 
For these reasons the constraints from \cite{Sirunyan:2018ubx} are weaker for $M_1\sim \mu$,
but stronger for $M_1\gg \mu$.

Additional allowed decays are $\chi^0_{2,3} \to H_{1} + \chi^0_1$ and
$\chi^0_{2,3} \to A_1 + \chi^0_1$ with mostly singlet-like and possibly very light $H_1$
and $A_1$. The corresponding Yukawa couplings can originate from the term $\frac{\kappa}{3} 
\hat S^3$ in the superpotential and singlino components of $\chi^0_{2,3}$, or from
the term $\lambda \hat S \hat H_u\cdot \hat H_d$ in the superpotential and higgsino
components of $\chi^0_1$. However, both Yukawa couplings relevant for decays into $H_1,\ A_1$ are
considerably smaller than the ones for decays into $Z$ or $H_{SM}$, and the branching
fractions are negligible unless on-shell decays into $Z$ (and hence into $H_{SM}$) are
kinematically impossible. In these latter cases the $BR(\chi^0_{2} \to H_1 + \chi^0_1)$
can become $\sim 100\%$. 

Another exception are light binos with masses below the ones of higgsinos
(which are now $\chi^0_{3,4}$). Then, if kinematically allowed, decays $\chi^0_{3,4} \to H_1 + \chi^0_2$
can have sizeable branching fractions. These imply more involved decay cascades of
$\chi^0_{3,4}$ which we consider (conservatively) not to contribute to the signals studied in
\cite{Sirunyan:2018ubx}.

On the other hand the decays into both $Z$ and $H_{SM}$ dominate
in most of the parameter space consistent with constraints from dark matter, and generally
both decays have similar branching fractions of ${\cal O}(30-70\%)$.

In the realistic light higgsino-singlino scenario of the NMSSM considered here,
the mostly higgsino-like fermions $\chi^\pm_1$, $\chi^0_2$ and $\chi^0_3$ are
not exactly degenerate due to mixing, but the masses satisfy typically
$M_{\chi^0_3}\gsim M_{\chi^\pm_1} \gsim M_{\chi^0_2}$ with
$M_{\chi^0_3}-M_{\chi^\pm_1} \sim M_{\chi^\pm_1}-M_{\chi^0_2} \lsim {\cal O}(20$~GeV).
Mixing affects also their production cross sections and branching
fractions, the latter are computed using the code {\sf NMSDECAY}~\cite{Das:2011dg}
(based on {\sf SDECAY}~\cite{Muhlleitner:2003vg}).
In order to obtain limits on these realistic scenarios into the considered final
states we proceed as follows:

First we consider separately the pairs $P_1=\chi^\pm_1$, $\chi^0_2$ (with typically
$M_{\chi^0_2} <  M_{\chi^\pm_1}$) and $P_2=\chi^\pm_1$, $\chi^0_3$ (with typically
$M_{\chi^0_3} >  M_{\chi^\pm_1}$). Using Prospino 2 at NLO \cite{Beenakker:1999xh}
we compute the production cross section at 13~TeV assuming pure higgsinos. For both
$P_1$ and $P_2$, the production cross sections are weighted by $X_1/(X_1+X_2)$, 
$X_2/(X_1+X_2)$, respectively, where $X_i$ are the products of the corresponding
couplings squared of $\chi^0_i,\ \chi^\pm_1$ to $W^{\pm}$ (relevant for the production
cross section) and the branching fractions $BR(\chi^0_i \to Z^{(*)} + \chi^0_1)$.
(For a bino mass $M_1 \lsim \mu$, $\chi^0_2$ and $\chi^0_3$ should be
replaced by the mostly higgsino-like neutralinos here and below.)

Decays into $H_{SM}$, $H_1$ and $A_1$, on which the limits are much weaker, are not taken into
account for estimates of the signal strength. Hence our limits will be conservative.

Next we look for a triplet of degenerate higgsinos with common mass $M_h$ which
would have the same production cross section times branching fraction.
The contributions of $P_1$ and
$P_2$ to the effective production cross section times branching fraction
of the ``fictitious'' degenerate triplet are considered according to their
relative weights $X_i/(X_1+X_2)$. $M_h$
is found from a table as function of $M_{\chi^0_3}-M_{\chi^\pm_1}$ and
$M_{\chi^\pm_1}-M_{\chi^0_2}$ constructed again with help of Prospino.
Typically one finds $M_h \sim M_{\chi^\pm_1}$.
 
Finally, the effective production cross section times branching fraction
of the ``fictitious'' degenerate triplet is rescaled (mildly) by the ratio
of higgsino production cross section from \cite{HiggsinoXS} with respect to
Prospino~2, and compared to the upper bounds
in the data files corresponding to Figs.~7 and 8a in \cite{Sirunyan:2018ubx},
whichever is stronger.

\section{Results}

We parametrize the higgsino-singlino scenario by $M_{\chi^0_1}$ and $M_{\chi^\pm_1}\approx
\mu$. In Figure~2 we show which regions are excluded by the combined
constraints from the dark matter relic density, limits on spin-dependent and
spin-independent dark matter direct detection cross sections, and CMS
\cite{Sirunyan:2018ubx}, for arbitrary bino mass $M_1$ in red.
Assuming $M_1\gsim 300$~GeV, the blue regions are excluded in addition.
(The structures for
$M_{\chi^0_1}\sim 5-60$~GeV and $M_{\chi^\pm_1}\sim 190-270$~GeV originate from
corresponding structures in the data files corresponding to Figs.~7 and 8a in 
\cite{Sirunyan:2018ubx}.)

\begin{figure}
\begin{center}
\includegraphics[width=12cm]{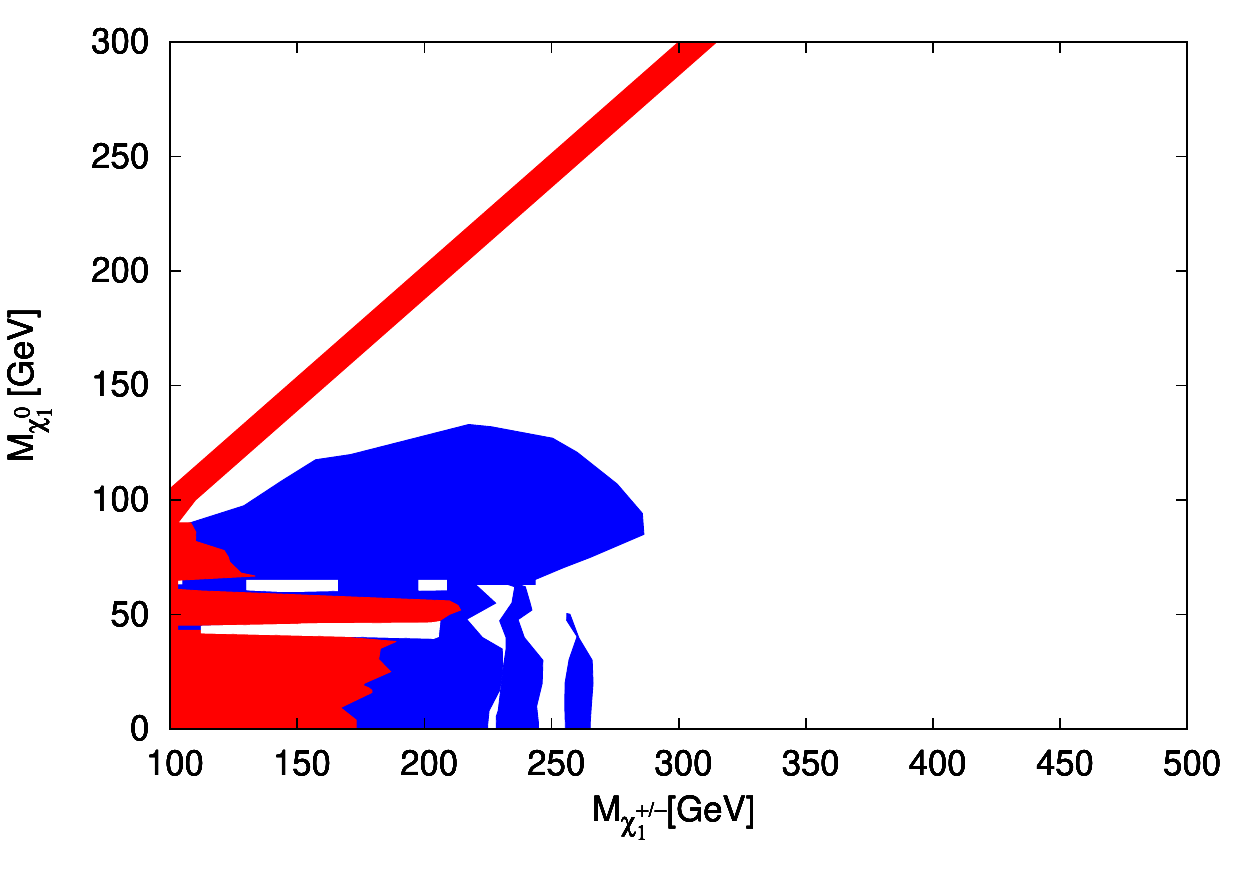}
\end{center}
\caption{Regions in the pNMSSM with heavy squarks, sleptons and gauginos
which are excluded by the combined
constraints from the dark matter relic density, limits on spin-dependent and
spin-independent dark matter direct detection cross sections, and searches by CMS
\cite{Sirunyan:2018ubx}. Red: arbitrary bino mass $M_1 \sim \mu$ or $M_1 < \mu$.
Blue: $M_1=300$~GeV.}
\end{figure}

The following remarks are in order: First, in the diagonal red band for
$M_{\chi^0_1}\lsim M_{\chi^\pm_1}$ the LSP has a large higgsino component
through mixing, and the relic density is reduced below the WMAP/Planck
value via coannihilation. In the blue/red ``bulk'' for $M_{\chi^0_1}\gsim 65$~GeV the combined
constraints from the relic density and direct detection cannot be
satisfied simultaneously. The constraints from CMS are not important there.

For $M_{\chi^0_1}\lsim 65$~GeV the constraints from CMS start to play the dominant
r\^ole and exclude regions notably for singlinos below
the $Z$ and $H_{SM}$ funnels for dark matter annihilation. As discussed
above these constraints from CMS depend on the assumptions on $M_1$, and are stronger
for $M_1 \gsim 300$~GeV. 

Around the $Z$-funnel ($M_{\chi^0_1}\sim 45$~GeV) and the
$H_{SM}$ funnel ($M_{\chi^0_1}\sim 62$~GeV) the singlet-like (pseudo-)scalars $H_1/A_1$ are not
needed for dark matter annihilation. The resulting freedom in the NMSSM parameter space
allows them to be light and to have sizeable couplings to the neutral higgsinos
$\chi^0_{2,3}$. This allows for large branching fractions for $\chi^0_{2,3}\to \chi^0_1 +H_1/A_1$
which circumvent the constraints from searches for $\chi^0_{2,3} \to Z/H_{SM} + \chi^0_1$.

Hence, both due to the reduced production cross sections for
higgsino-like charginos/neu\-tra\-linos and reduced branching fractions into
the $W/Z$ final states, the excluded regions are considerably smaller
than for simplified models in \cite{Aaboud:2017leg,Aaboud:2018jiw,
Sirunyan:2018ubx,Sirunyan:2017lae} assuming wino-like charginos/neutralinos.

Next we consider the NUH-NMSSM.
In Figure~3 we show possible points in the $M_{\chi^0_1}-M_{\chi^\pm_1}$
plane satisfying the combined constraints from the dark matter relic density,
dark matter direct detection and CMS as before. In addition we show the
necessary fine tuning among the parameters at the GUT scale. The measure for
fine tuning is the one of Barbieri and Giudice \cite{Barbieri:1987fn}
implemented in {\sf NMSSMTools} following \cite{Ellwanger:2011mu}. We are aware
of the fact that the measure for fine tuning in \cite{Barbieri:1987fn} has to
be taken with care and might sometimes be misleading (too strong), but it
serves nevertheless as a rough handwaving guide. For a given couple
$M_{\chi^0_1}-M_{\chi^\pm_1}$ the fine tuning is not unique but depends
also on other parameters; the points selected for Figure~3 (and others
below) correspond to the minimal possible fine tuning within bins of
size $1~\text{GeV}\times 1~\text{GeV}$.

\begin{figure}
\begin{center}
\includegraphics[width=11cm,trim=0 0.5cm 0 1.5cm, clip]{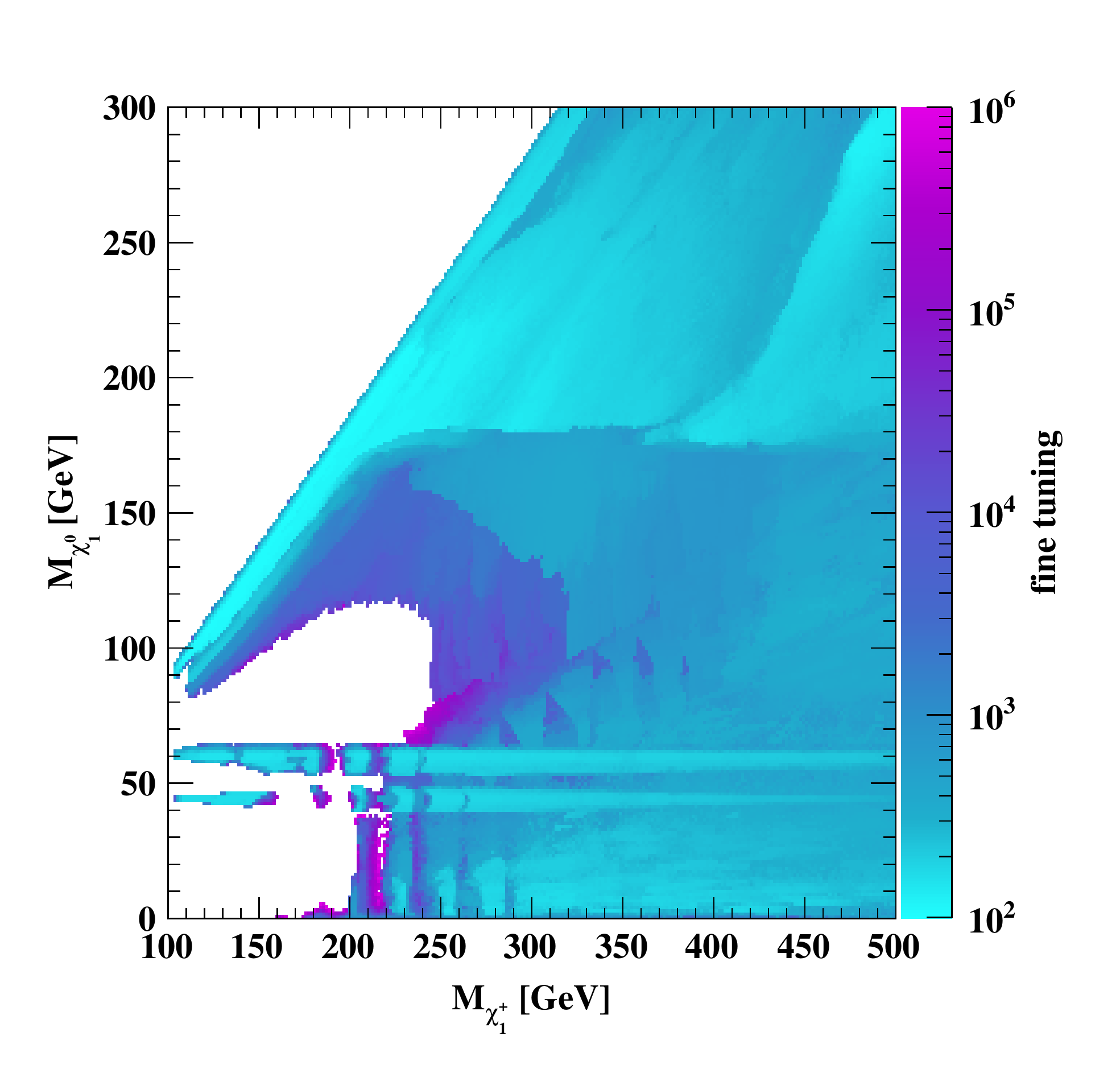}
\end{center}
\caption{Points with minimal fine tuning in the NUH-NMSSM satisfying the
combined constraints from the dark matter relic density,
dark matter direct detection and CMS.}
\end{figure}

One finds that the complete unexcluded white (and parts of the blue) region
in Fig.~2 is also allowed in the NUH-NMSSM. Regions where the fine tuning
is relatively weak (not far above $10^2$, i.e. light blue) correspond to
different LSP annihilation processes:

\noi i) Along $M_{\chi^0_1} \lsim M_{\chi^\pm_1}$: co-annihilation with higgsinos,
and/or annihilation among higgsinos before these decay into the nearly degenerate
$\chi^0_1$;

\noi ii) via $Z$- or $H_{SM}$ funnels for $M_{\chi^0_1} \approx 45$~GeV or
$M_{\chi^0_1} \approx 62$~GeV;

\noi iii) via a pseudoscalar in the s~channel if $M_{\chi^0_1} \lsim 30$~GeV
and $M_{\chi^\pm_1}\gsim 200$~GeV;

\noi iv) via both a pseudoscalar and a far off-shell $Z\to t\bar{t}$ for
$M_{\chi^0_1} \gsim 175$~GeV.

Annihilation via a pseudoscalar in the s~channel is also typical in the other
regions. Large fine tuning is required notably in regions where the constraints
from trilepton searches at the LHC are strong (recall that these are not continuous
in $M_{\chi^\pm_1}$); then large radiative corrections to the Higgs sector (i.e.
large stop masses) are necessary in order to satisfy simultaneously the constraints
from direct detection and the relic density.

The different dark matter annihilation processes imply different dark matter
detection cross sections. On the left hand side of Figure~4 we show the
spin-independent dark matter-proton cross section $\sigma_{pSI}$ for points with
minimal fine tuning. In the region i) where co-annihilation with higgsinos and/or annihilation among
higgsinos reduces the $\chi^0_1$ relic density to the observed value, $\chi^0_1$
can be very singlino-like implying a small $\sigma_{pSI}$. A very singlino-like
$\chi^0_1$ corresponds to small higgsino-singlino mixing or a small coupling
$\lambda$, see the right hand side of Figure~4. Also annihilation of a light $\chi^0_1$
($M_{\chi^0_1} \lsim 20$~GeV) via a light pseudoscalar in the s~channel (or of a
heavier $\chi^0_1$ via the $H_{SM}$ or $Z$ funnels)
can correspond to a mostly singlino-like $\chi^0_1$ in order to
avoid a too small relic density, implying $\lambda \lsim 0.2$. In most of the
other regions where annihilation proceeds via a pseudoscalar in the s~channel
one finds $\lambda \gsim 0.2$, i.e. not too small according to the arguments given
below eq.~(3.1). 

$\sigma_{pSI}$ can even fall below the expected background from
neutrinos \cite{Billard:2013qya}, see Figure~5
(a phenomenon observed before in \cite{Ellwanger:2014dfa,Badziak:2015exr,Beskidt:2017xsd}).
This can happen for $M_{\chi^0_1} \lsim 20$~GeV, and for larger $M_{\chi^0_1}$
in the region i). Such small cross sections can also occur for larger $\lambda$ (see
the dark blue regions/spots corresponding to $\sigma_{pSI}\lsim 10^{-13}$~pb
on the left hand side of Figure~4), due to negative interferences
among the t-channel exchanges of the SM-like Higgs boson and $H_1$.

\begin{figure}
\begin{center}
\includegraphics[width=8cm,trim=0cm 0.5cm 0cm 1.5cm, clip]{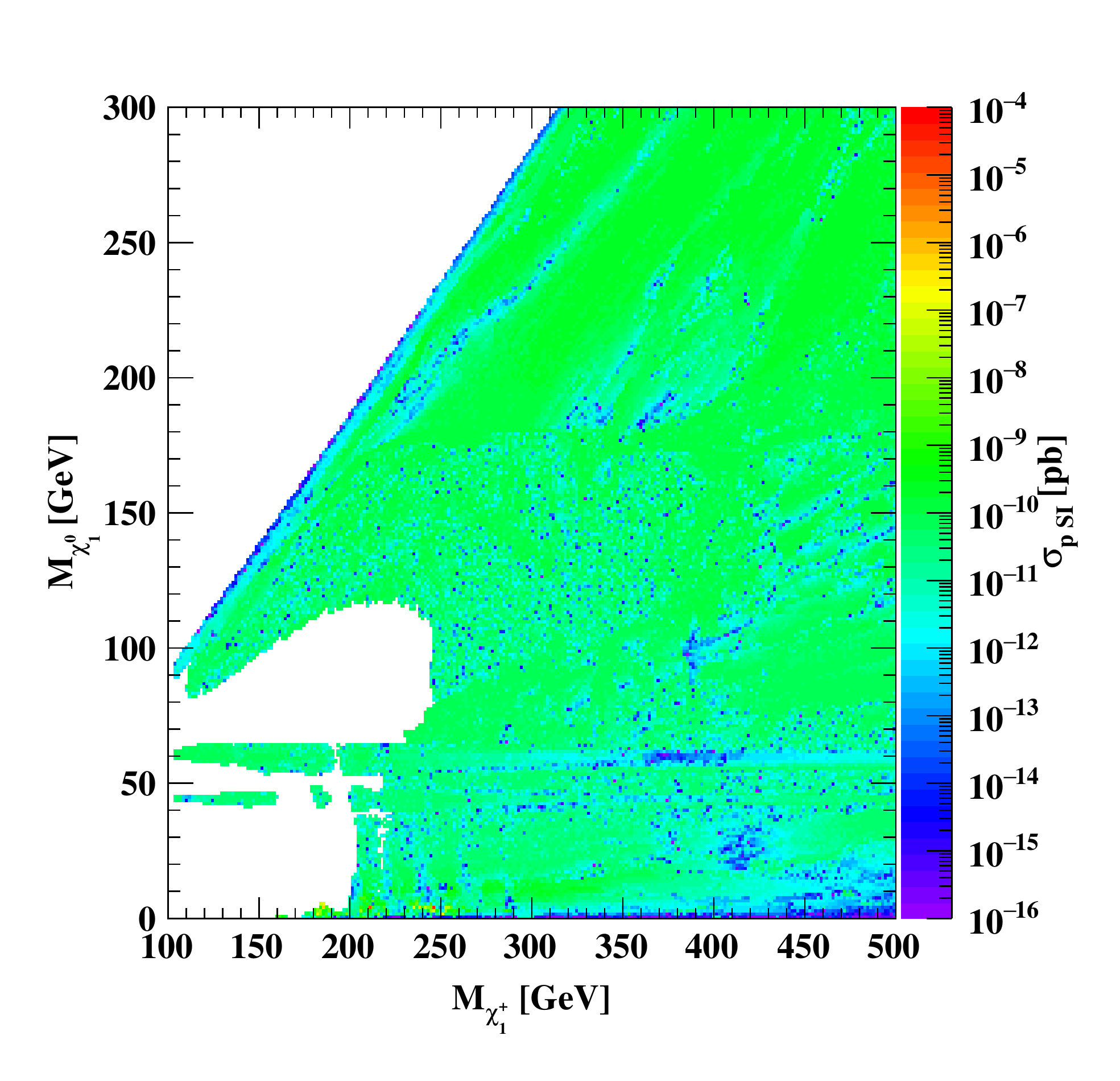}
\includegraphics[width=8cm,trim=0cm 0.5cm 0cm 1.5cm, clip]{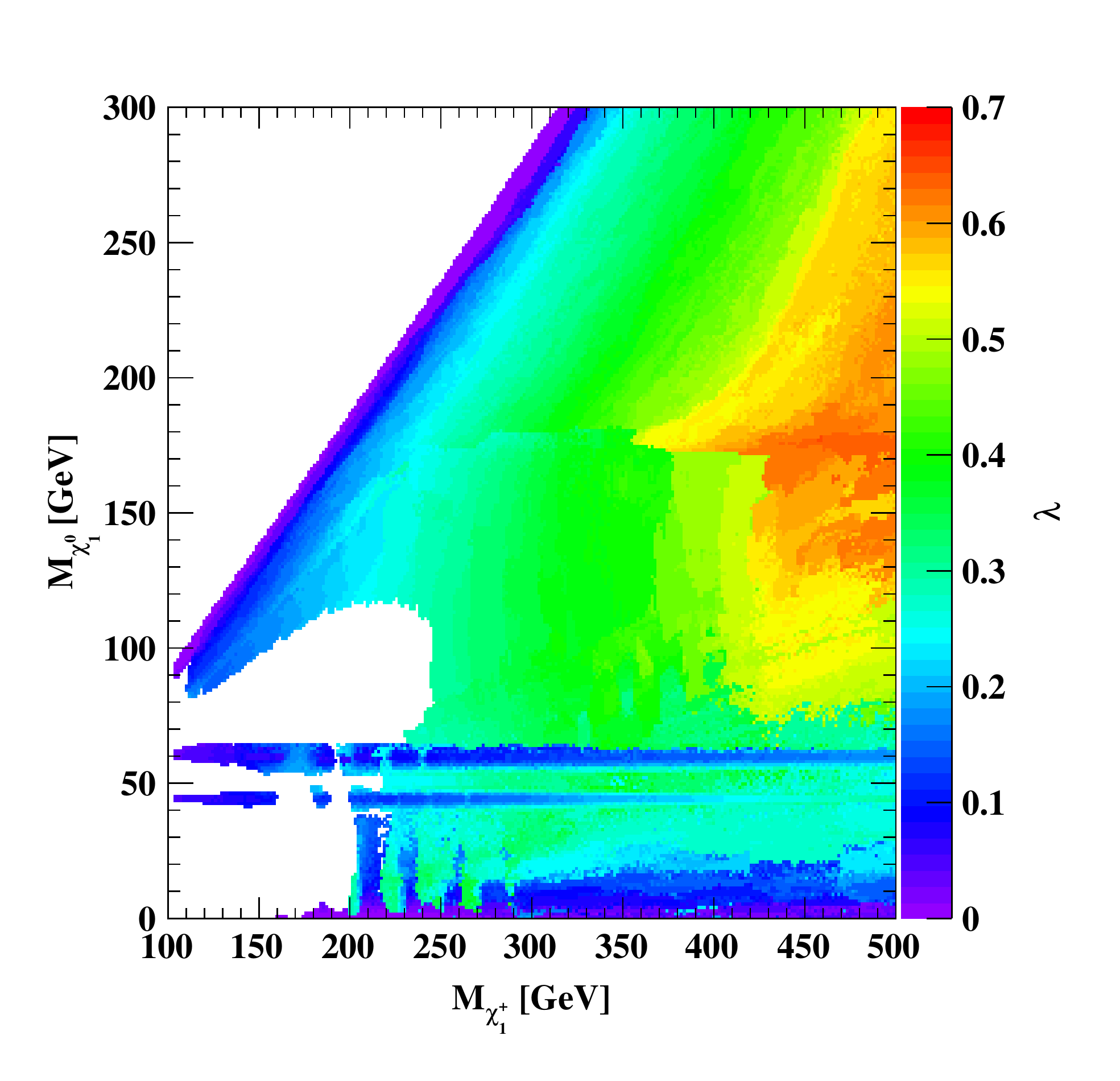}
\end{center}
\caption{Left hand side: 
Spin-independent dark matter-proton cross section
$\sigma_{pSI}$ (in~pb) for points with minimal fine tuning. 
Right hand side:
The NMSSM specific coupling $\lambda$ for points with minimal fine tuning.
}
\end{figure}

\begin{figure}
\begin{center}
\includegraphics[width=13cm,trim=0 0.5cm 0 1cm, clip]{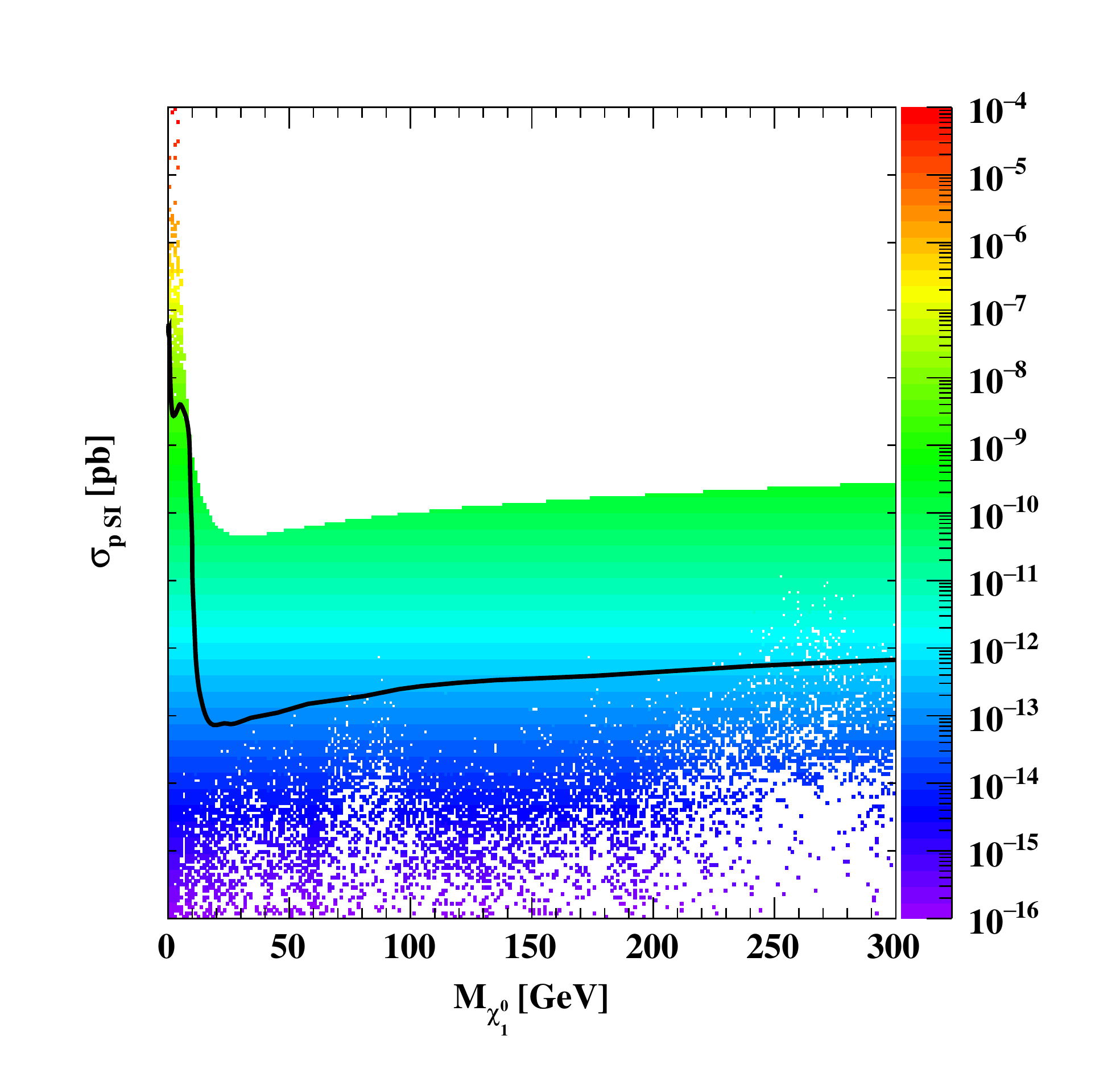}
\end{center}
\caption{$\sigma_{pSI}$ (in~pb) as function of $M_{\chi^0_1}$
 for points with minimal fine tuning together with the
expected background from neutrinos \cite{Billard:2013qya} as a black line.}
\end{figure}

\section{Benchmark points and planes}

\subsection{$W + Z/H_{SM}$ final states}

As we have seen the searches for $pp\to \chi^\pm + \chi^0_i$ with
$\chi^\pm \to W^{\pm} + \chi^0_1$,
$\chi^0_i \to Z + \chi^0_1$ or $\chi^0_i \to H_{SM} + \chi^0_1$
cover also regions of the parameter space of the light higgsino-singlino sector. 

We recall that the realistic scenarios within the NMSSM differ from the
simplified models used for current interpretations of limits on cross sections times branching
fractions as follows:

\begin{itemize}
\item Two neutral higgsino-like neutralinos can be produced together with a
higgsino-like chargino. These three states are not exactly degenerate due to mixing.
Mixing with the singlino (and/or the bino) also reduces their production cross
sections relative to pure higgsino-like states.

It would be desirable if the experimental collaborations could check the uncertainty
introduced by the replacement of a degenerate chargino-higgsino system by
a non-degenerate one using a weighting according to production cross sections and branching
fractions as above. If these uncertainties are not too large, benchmark planes
employing degenerate chargino-neutralino systems can simulate realistic scenarios
within the NMSSM with reasonable accuracy.
(Non-degenerate higgs\-inos have been considered by ATLAS in
\cite{Aaboud:2017leg}. However, there the lighter higgsino was assumed to be
the LSP.)


\item Branching fractions are different. Whereas the $BR(\chi^\pm_1\to \chi^0_1 + W^{\pm (*)})$
is (nearly) always 100\%, the ones of the two neutral higgsino-like neutralinos can
vary over wide ranges; examples are given for benchmark points in Table~1.
 All points satisfy constraints from the dark
matter relic density and direct dark matter detection.
(Branching fractions into $A_1$ are always below 4\% and omitted for simplicity.)

\end{itemize}

\begin{table}[ht]
\center
\begin{tabular}{|c|c|c|c|c|c|c|}
\hline
                                    & P1   & P2   & P3   & P4   & P5   & P6   \\ \hline
$M_{\chi^\pm_1}$                    & 265  & 261  & 219  & 286  & 276  & 193  \\ \hline 
$M_{\chi^0_1}$                      & 3.2  & 40   & 62   & 85   & 107  & 150  \\ \hline
$M_{\chi^0_2}$                      & 250  & 244  & 206  & 261  & 257  & 197  \\ \hline
$M_{\chi^0_3}$                      & 285  & 278  & 236  & 306  & 293  & 205  \\ \hline
$M_{H_1}$                           & 56   & 35   & 59   & 20   & 3    & 60   \\ \hline
$M_{A_1}$                           & 76   & 78   & 63   & 167  & 205  & 259  \\ \hline
\hline
$BR(\chi^0_2\to \chi^0_1 + Z)$      & 0.40 & 0.30 & 0.84 & 0.73 & 0.13 & 0.95* \\ \hline
$BR(\chi^0_2\to \chi^0_1 + H_{SM})$ & 0.48 & 0.64 & 0.09 & 0.22 & 0.77 & 0.00 \\ \hline
$BR(\chi^0_2\to \chi^0_1 + H_{1})$  & 0.08 & 0.05 & 0.02 & 0.03 & 0.10 & 0.00 \\ \hline
$BR(\chi^0_3\to \chi^0_1 + Z)$      & 0.57 & 0.70 & 0.39 & 0.34 & 0.89 & 0.99*\\ \hline
$BR(\chi^0_3\to \chi^0_1 + H_{SM})$ & 0.33 & 0.24 & 0.56 & 0.61 & 0.09 & 0.00 \\ \hline
$BR(\chi^0_3\to \chi^0_1 + H_{1})$  & 0.06 & 0.02 & 0.03 & 0.05 & 0.02 & 0.00 \\ \hline
\hline
Xsect $\to \chi^\pm_1+\chi^0_2$ [fb]& 125  & 139  & 318  &  85  &  93  & 295  \\ \hline
Xsect $\to \chi^\pm_1+\chi^0_3$ [fb]& 128  & 141  & 258  &  96  & 115  & 437  \\ \hline
\end{tabular}
\caption{Masses (in GeV) and branching fractions of benchmark points of the pNMSSM. Branching fractions
into $Z$ with a star indicate off-shell decays. The production cross
sections in the last two lines are obtained by prospino~2 at NLO \cite{Beenakker:1999xh}.}
\label{t1}
\end{table}

For all points P1 -- P6 the sums of the branching fractions
$BR(\chi^0_{2,3} \to Z + \chi^0_1)$ and $BR(\chi^0_{2,3} \to H_{SM} + \chi^0_1)$
are close to or above 90\%. The individual branching fractions of
$\chi^0_{2,3}$ vary considerably; the {\it average} branching fractions are,
however, quite stable: $50\%-70\%$ into $Z + \chi^0_1$, $50\%-30\%$ into $H_{SM} + \chi^0_1$.
Even for P6, where only off-shell decays of $\chi^0_{2,3}\to X + \chi^0_1$
are possible, one finds $BR(\chi^0_{2} \to Z^* + \chi^0_1)\sim 100\%$.
P1 -- P3 correspond to a relatively light $\chi^0_{1}$. Using the averaging described
above we find that they are not far from being excluded, hence they may serve
to test the averaging described above.

\vspace*{2mm}

{\bf Benchmark planes:}
In terms of a single $\chi^0_{2}$ representing the average branching fractions
of $\chi^0_{2}$ and $\chi^0_{3}$ and assuming $M_{\chi^\pm_1} = M_{\chi^0_2} = M_{\chi^0_3}$,
it remains useful to study upper limits on production cross sections in the
plane $M_{\chi^\pm_1} - M_{\chi^0_1}$. For the branching fractions of $\chi^0_{2}$
it is reasonable to assume
$BR(\chi^0_{2} \to Z + \chi^0_1)=50\%$, $BR(\chi^0_{2} \to H_{SM} + \chi^0_1)=50\%$
(as already done in  Fig. 8c in \cite{Sirunyan:2018ubx}) or
$BR(\chi^0_{2} \to Z + \chi^0_1)=70\%$, $BR(\chi^0_{2} \to H_{SM} + \chi^0_1)=30\%$.

\subsection{$W + H_1/A_1$ final states}

\begin{table}
\center
\begin{tabular}{|c|c|c|c|c|c|c|}
\hline
                                    & P7   & P8   & P9  & P10 & P11 & P12 \\ \hline
$M_{\chi^\pm_1}$                    & 129  & 237  & 118 & 158 & 210 & 226 \\ \hline 
$M_{\chi^0_1}$                      & 97   & 160  & 45  & 47  & 50  & 60  \\ \hline
$M_{\chi^0_2}$                      & 131  & 238  & 110 & 123 & 128 & 180 \\ \hline
$M_{\chi^0_3}$                      & 140  & 248  & 128 & 172 & 222 & 240 \\ \hline
$M_{\chi^0_4}$                      & 303  & 355  & 302 & 183 & 224 & 246 \\ \hline
$M_{H_1}$                           & 32   & 25   & 35  &  43 & 5   & 62  \\ \hline
$M_{A_1}$                           & 174  & 290  & 42  &  37 & 49  & 21  \\ \hline 
\hline
$BR(\chi^0_2\to \chi^0_1 + Z)$      & 0.00 & 0.00 & 0.10*& 0.02*& 0.00 & 0.16 \\ \hline
$BR(\chi^0_2\to \chi^0_1 + H_{SM})$ & 0.00 & 0.00 & 0.00 & 0.00 & 0.00 & 0.00 \\ \hline
$BR(\chi^0_2\to \chi^0_1 + H_{1})$  & 1.00 & 1.00 & 0.38 & 0.27 & 1.00 & 0.01 \\ \hline
$BR(\chi^0_2\to \chi^0_1 + A_{1})$  & 0.00 & 0.00 & 0.52 & 0.71 & 0.00 & 0.02 \\ \hline 
$BR(\chi^0_2\to \nu_\tau + \tilde{\nu}_\tau)$  
                                    & 0.00 & 0.00 & 0.00 & 0.00 & 0.00 & 0.81 \\ \hline 
\hline
$BR(\chi^0_3\to \chi^0_1 + Z)$      & 0.96*& 0.88*& 0.33*& 0.80 & 0.25 & 0.36 \\ \hline
$BR(\chi^0_3\to \chi^0_1 + H_{SM})$ & 0.00 & 0.00 & 0.00 & 0.09 & 0.39 & 0.39 \\ \hline
$BR(\chi^0_3\to \chi^0_1 + H_{1})$  & 0.04 & 0.12 & 0.61 & 0.08 & 0.07 & 0.02 \\ \hline
$BR(\chi^0_3\to \chi^0_1 + A_{1})$  & 0.00 & 0.00 & 0.03 & 0.01 & 0.00 & 0.00 \\ \hline
$BR(\chi^0_3\to \chi^0_2 + Z)$      & 0.00 & 0.00 & 0.03*& 0.00 & 0.06 & 0.00 \\ \hline
$BR(\chi^0_3\to \chi^0_2 + H_1)$    & 0.00 & 0.00 & 0.00 & 0.02 & 0.23 & 0.00 \\ \hline 
$BR(\chi^0_3\to \tau^\pm + \tilde{\tau}^\mp)$  
                                    & 0.00 & 0.00 & 0.00 & 0.00 & 0.00 & 0.17 \\ \hline 
$BR(\chi^0_3\to \nu_\tau + \tilde{\nu}_\tau)$  
                                    & 0.00 & 0.00 & 0.00 & 0.00 & 0.00 & 0.06 \\ \hline 
\hline
$BR(\chi^0_4\to \chi^0_1 + Z)$      &      &      &      & 0.44 & 0.86 & 0.23 \\ \hline
$BR(\chi^0_4\to \chi^0_1 + H_{SM})$ &      &      &      & 0.01 & 0.06 & 0.03 \\ \hline
$BR(\chi^0_4\to \chi^0_1 + H_{1})$  &      &      &      & 0.01 & 0.02 & 0.00 \\ \hline
$BR(\chi^0_4\to \chi^0_1 + A_{1})$  &      &      &      & 0.00 & 0.02 & 0.00 \\ \hline
$BR(\chi^0_4\to \chi^0_2 + Z)$      &      &      &      & 0.00 & 0.04 & 0.00 \\ \hline
$BR(\chi^0_4\to \chi^0_2 + H_1)$    &      &      &      & 0.51 & 0.00 & 0.07 \\ \hline
$BR(\chi^0_4\to \tau^\pm + \tilde{\tau}^\mp)$  
                                    &      &      &      & 0.00 & 0.00 & 0.56 \\ \hline 
$BR(\chi^0_4\to \nu_\tau + \tilde{\nu}_\tau)$  
                                    &      &      &      & 0.00 & 0.00 & 0.10 \\ \hline 
\hline
Xsect $\to \chi^\pm_1+\chi^0_2$ [fb]& 1319 & 186  & 3138 & 670  &  78  & 145  \\ \hline
Xsect $\to \chi^\pm_1+\chi^0_3$ [fb]& 1759 & 212  & 2376 & 829  & 295  & 241  \\ \hline
Xsect $\to \chi^\pm_1+\chi^0_4$ [fb]& 9    &   7  &   8  & 437  & 316  & 164  \\ \hline
\end{tabular}
\caption{Masses (in GeV) and branching fractions of benchmark points of the pNMSSM.
Branching fractions
into $Z$ with a star indicate off-shell decays. The production cross
sections in the last three lines are obtained by prospino~2 at NLO \cite{Beenakker:1999xh}.
Branching fractions of $\chi^0_{4}$ are ignored if their production rate is
negligible. P12~has $M_{\tilde{\tau}_1} \sim 178$~GeV, $M_{\tilde{\nu}_\tau} \sim 162$~GeV.}
\label{t2}
\end{table}

If the mostly singlet-like scalars $H_1$ and/or pseudoscalars $A_1$ are light and
$M_{\chi^0_{2,3}} - M_{\chi^0_{1}} < M_Z$, the branching fractions
$BR(\chi^0_{2,3}\to \chi^0_1 + H_1/A_1)$ can become dominant. The points P7 -- P9 in
Table~2 correspond to such cases.

For P7 and P8, only the lighter higgsino $\chi^0_{2}$ decays via $H_1$, whereas the
heavier higgsino $\chi^0_{2}$ still prefers decays via an off-shell $Z^*$.
For P9 decays via $H_1$ or $A_1$ are dominant for both higgsinos. Although
$H_1$ and $A_1$ have $\sim 90\%$ branching fractions into $b\bar{b}$ the fact
that their (generally different) masses are not known a priori will
make it hard to detect such scenarios despite possibly large production cross
section like for P7 and P9. Note that $\chi^\pm_{1}$ will decay via an off-shell~$W^*$.

\vspace*{2mm}

{\bf Benchmark planes:}
Still we propose studies in the plane $M_{\chi^\pm_1} - M_{\chi^0_1} < M_Z$ for
such scenarios. Instead of $W + Z/H_{SM}$ final states the aims are now
$W^{(*)} + Z/H_1$ or pure $W^{(*)} + H_1$ final states. Dedicated studies could
start with different assumptions on $M_{H_1}$, and optimize cuts correspondingly.
Such studies seem necessary in order to all scenarios of light higgsino decays
in the NMSSM.


\subsection{Light binos and staus}

For the previous points P1 -- P9 the bino mass parameter $M_1$ satisfies $M_1 > \mu$,
and the production of the bino-like $\chi^0_4$ plays little role, see the cross
sections in the last line of Table~2. The sitation changes if $M_1 \lsim \mu$ and
$\chi^0_2$ is mostly bino-like, but $\chi^0_{3,4}$ are mostly higgsino-like. Now,
due to mixing, all $\chi^0_{2,3,4}$ can have sizeable production cross sections (still
together with $\chi^\pm_1$), see P10 -- P12 in Table~2.

Moreover new decay cascades become possible. An example is P10 where $\chi^0_{4}$
decays dominantly into $\chi^0_{2}+ H_1$, and subsequently $\chi^0_{2}$
decays dominantly into $\chi^0_{1}+ A_1$. This point is thus both interesting and
challenging.

Finally staus $\tilde{\tau}$ (scalar partners of a right- or left handed tau or
tau-neutrino) can be lighter than the higgsinos. Then both 
charged and neutral higgsinos can have large branching fractions into these
states. 

Decays of (degenerate) charginos/neutralinos into staus have been considered
by CMS in \cite{CMS:2017wox}. Limits on production cross sections as function of
$M_{\chi^\pm}$ and $M_{\chi_1^0}$ depend on the assumed stau mass.
The derived limits in the $M_{\chi^\pm}-M_{\chi_1^0}$ plane
can be quite strong for $M_{\tilde{\tau}}$ near $M_{\chi^\pm}$, but
production cross sections for wino-like charginos/neutra\-linos are assumed.
For higgsino-like charginos/neutra\-linos as considered here these limits
will be weaker.

An example is given by P12 in Table~2 for which the stau masses are
$M_{\tilde{\tau}_1} \sim 178$~GeV, $M_{\tilde{\nu}_\tau} \sim 162$~GeV.
Here $\chi^0_{2}$ is also dominantly bino-like. One finds quite different
decays for the three neutralinos $\chi^0_{2,3,4}$, but the branching fractions
into $\tilde{\tau}^\pm$ and/or $\tilde{\nu}_\tau$ are sizeable. Not shown
in Table~2 are the branching fractions of $\chi^\pm_1$ which are now
$BR(\chi^\pm_1\to \tilde{\nu}_\tau + \tau^\pm)\sim 34\%$,
$BR(\chi^\pm_1\to \tilde{\tau}_1^\pm + \nu_\tau)\sim 3\%$ with 63\%
remaining for $\chi^\pm_1\to \chi^0_1 + W^\pm$.

\section{Summary}

Light singlinos in the NMSSM are still promising candidates for dark matter,
and light higgsinos (a $\mu$ parameter not too far above the electroweak scale)
are natural. Since the direct detection cross sections for singlino-like dark
matter may fall below the neutrino floor, searches for this scenario at
colliders are particularly relevant.
The first purpose of the present paper is to interprete recent searches for electroweak
production of supersymmetric particles at the LHC in this scenario, combined
with constraints from dark matter.

Given the extended parameter space and the extended neutralino sector of the NMSSM
implying wide ranges of masses, mixing angles and branching fractions,
this is not a simple task. Here we assume not only that the singlino has all
required properties of a good dark matter candidate, but also that winos
have masses $\gsim 600$~GeV which is motivated by
wino-gluino mass unification at the GUT scale and lower bounds on the gluino mass.
(Since wino decay cascades would be lengthy and very different from simplified models,
constraints on wino pair production cannot be applied here. Allowing for lighter
winos would require separate analyses.)

Within these assumptions our derived limits can be considered as quite conservative
since we did not combine the $W/Z$ final state with other search channels or other
higgsino pair production modes. Still we find that the limits from CMS in
\cite{Sirunyan:2018ubx} exclude 
regions of $M_{\chi^\pm_1}\sim \mu$ which would be allowed by constraints on
dark matter alone. Assuming a bino mass above $\sim 300$~GeV, motivated again by
gaugino mass unification at the GUT scale and lower bounds on the gluino mass,
few regions with $M_{\chi^\pm_1}\lsim 250$~GeV remain viable for a singlino mass below
$\sim 60$~GeV.

However, we have also identified scenarios which are not visible
in the search for $W + Z/H_{SM}$ final states: Neutralino decays into NMSSM
specific light scalars or pseudoscalars instead of $Z/H_{SM}$, mixed
bino-higgsino scenarios leading to possibly more involved decay chains,
and notably staus lighter than higgsinos. All these scenarios are favoured
by the good properties of the singlino LSP as dark matter. Since
these different scenarios would have different impacts on all supersymmetric particle decay
cascades their studies merit considerable efforts.

To this end we propose benchmark points and planes. 
At least once different signal regions are combined using ranges of
branching fractions discussed in Section~5 we are convinced that
future searches at the LHC can test considerably more promising regions in the
singlino-higgsino mass plane.
In general the consequences of these unconventional scenarios are essentially
that the sums of the branching fractions into $W + Z/H_{SM}$ final states
do not add up to 100\%, i.e. upper limits on these branching fractions as
function of the involved masses will be insufficient sources of information.

\section*{Acknowledgements}

We thank Lei Wu for helpful comments.
The authors acknowledge the support of France-Grilles and the OCEVU Labex
(ANR-11-LABX-0060) for providing computing resources on the French National
Grid Infrastructure, and support from the French research project D\'efi InFIniti - AAP 2017. 
U.~E. acknowledges support from the European Union's Horizon 2020
research and innovation programmes H2020-MSCA-RISE No. 645722
(NonMinimalHiggs) and under the Marie Sklodowska-Curie grant agreement No 690575
(InvisiblesPlus).


\end{document}